\documentclass[prb,amsmath,preprint]{revtex4}
\usepackage{graphicx}
\usepackage{amssymb}

\pdfoutput=1
\begin{document}

\title{Theory of Transient Excited State Absorptions in Solid Pentacene
with Implications for Singlet Fission}
\author{Souratosh Khan}
\affiliation{Department of Physics, University of Arizona
Tucson, AZ 85721}
\author{Sumit Mazumdar}
\affiliation{Department of Physics, University of Arizona}
\affiliation{Department of Chemistry and Biochemistry, University of Arizona}
\affiliation{College of Optical Sciences, University of Arizona}
\date{\today}
\begin{abstract}
\noindent {\bf We report the first theoretical calculations of excited state absorptions
(ESAs) from the singlet and triplet excitons, as well as the key intermediate in the singlet fission (SF) process,
the spin singlet multiexciton triplet-triplet state, for solid pentacene with herringbone crystal structure. Our goal
is to compare theoretical results against ultrafast transient photoinduced absorption (PA) measurements and their interpretations,
which have remained controversial.
We show that the elusive triplet-triplet state absorbs both in the visible and near infrared (NIR), at or close to the
PA energies assigned to the free triplet exciton. In contrast, the
triplet PA has nearly vanishing oscillator strength in the NIR within the rigid herringbone structure.
Observable oscillator strength for NIR triplet PA requires photoinduced enhancement of coupling between a pair of
neighboring pentacene molecules that confers significant charge-transfer (CT)
character to the triplet exciton. We discuss the implication of our results for efficient SF in pentacene and related materials.}
\end{abstract}
\maketitle
The photophysics of pentacene has been of strong interest for decades.
In recent years, it is being intensively studied as a material in which 
SF is nearly 100\% efficient \cite{Ehrler12a,Lee13a}. In principle each triplet exciton T$_1$ generated by SF can undergo charge dissociation
at the donor-acceptor interface of an organic heterostructure, doubling the photoconductivity expected from the dissociation
of the singlet exciton S$_1$. 
SF-driven enhancement of performance has been found in pentacene-C$_{60}$ solar cells\cite{Yoo04a,Congreve13a}
These observations
have led to a flurry of experimental \cite{Jundt95a,Thorsmolle09a,Marciniak09a,Chan11a,Wilson13b,Lee13a,Herz15a,Bakulin16a,Pensack16a,Walker13a} and theoretical \cite{Smith10a,Greyson10a,Kuhlman10a,Zimmerman11a,Smith13a,Beljonne13a,Chan13a,Zeng14a,Yost14a,Aryanpour15a} research on pentacene and its derivatives.

Although the thermodynamic requirement for SF, E(S$_1$) $\geq 2\times$ E(T$_1$), where E(...) is the energy of a state, is met widely,
relatively few organic molecular systems actually exhibit efficient SF.
Recent research has therefore focused on elucidating
the mechanism of SF, in particular, on determining the roles of the intermediates,
and that of materials morphology.
Experimental ultrafast transient absorption studies performed
on pentacene and 
bis(triisopropylsilylethynyl) (TIPS) pentacene,
in the solid state (polycrystalline films) \cite{Jundt95a,Thorsmolle09a,Wilson13b,Marciniak09a,Herz15a,Bakulin16a,Pensack16a} or as concentrated solutions \cite{Walker13a}.
have established that, 
(i) a triplet-triplet double excitation $^1$(TT)$_1$ is a key intermediate state in SF \cite{Swenberg,Smith10a,Smith13a,Burdett11a,Chan11a,Chan13a,Piland14a,Musser15b,Bakulin16a,Pensack16a} (here the superscript refers to spin multiplicity, and the subscript indicates that it is lowest triplet-triplet state), and
(ii) the $^1$(TT)$_1$ state consists of two spin triplet excitations on neighboring molecules, whose
spin angular momenta are quantum-entangled to give overall spin zero.
Beyond these, however, uncertainties remain.

Experimental detection of SF by time-resolved spectroscopy requires the observation of, 
(i) instantaneous pump-induced appearance of singlet S$_1 \to$ S$_N$ PA (here S$_N$ are higher singlet excitations)
(ii) paired decay of the singlet PA and the rise of triplet T$_1 \to$ T$_N$ PA which persists
 to long times. While these observations
were claimed 
in early measurements on pentacene films \cite{Jundt95a}, they are not considered conclusive as the measurements were in the visible range
of the electromagnetic spectrum, where singlet and
triplet PAs are overlapping.
More recent transient absorption studies \cite{Wilson13b,Walker13a} have assigned a broad PA at $\sim$ 750 nm$-$950 nm to triplet PA T$_1$ $\to$ T$_2$.
Although the growth of this NIR PA is concomitant with the decay of the singlet PA \cite{Wilson13b,Walker13a},
there are disagreements \cite{Smith10a,Smith13a} over this assignment. First, experimentally, the longest wavelength
at which monomer triplet PA is
found is at 505 nm \cite{Hellner72a}; this observation is in agreement with theoretical calculations \cite{Pabst08a,Chakraborty14a} that we have confirmed here. 
Second, it has been speculated \cite{Chan11a} that the PA in the NIR is from the
$^1$(TT)$_1$ rather than free T$_1$. Pensack {\it et al.} have recently claimed overlapping PAs from
T$_1$ and $^1$(TT)$_1$, in both NIR and the visible \cite{Pensack16a}.
Conversely, Herz {\it et al.}, based on 3-beam pump-depletion-probe
experiment, assign the NIR PA to both $^1$(TT)$_1$ and T$_1$,
and the visible PA at 500-550 nm to T$_1$ $\to$ T$_3$ alone \cite{Herz15a}. Clearly, the elucidation of the mechanism of SF in pentacene,
requires credible calculations of ESAs from all relevant states.

In the present work we report precise calculations of ESAs from S$_1$, T$_1$ and $^1$(TT)$_1$ in pentacene in the solid state.
Our results allow direct comparisons of theory against experiments, and lead to new insight to SF in
pentacene.
%
%
We have performed multiple reference singles and doubles configuration interaction (MRSDCI) \cite{Tavan87a,Chakraborty14a,Aryanpour15a} (see Methods)
calculations of ground and excited states of dimers of pentacene molecules in the herringbone lattice,
within the {\it extended} Pariser-Parr-Pople (PPP) Hamiltonian \cite{Aryanpour15a}.
The $\pi$-electron only PPP Hamiltonian \cite{Pariser53a,Pople53a} allows incorporating CI with up to quadruple excitations from the Hartree-Fock (HF) ground state of the pentacene dimer,
which is essential for obtaining precise theoretical description of the multiexciton $^1$(TT)$_1$ state \cite{Tavan87a}, and more importantly,
of ESA from the $^1$(TT)$_1$ to even higher energy states.
The extended PPP Hamiltonian is written as,

\begin{equation}
\label{PPP_Ham}
H_{\mathrm{PPP}}=H_{\mathrm{intra}}+H_{\mathrm{inter}}\,,
\end{equation}
\begin{equation}
\label{intra_Ham}
 H_{\mathrm{intra}}=\sum_{\mu,\langle ij \rangle,\sigma}t_{ij}^{\mu}(\hat{c}_{\mu i\sigma}^{\dagger}\hat{c}_{\mu j\sigma}^{}+\hat{c}_{\mu j\sigma}^\dagger \hat{c}_{\mu i\sigma}^{}) + U\sum_{\mu, i}\hat{n}_{\mu i\uparrow} \hat{n}_{\mu i\downarrow} + \sum_{\mu,i<j} V_{ij} (\hat{n}_{\mu i}-1)(\hat{n}_{\mu j}-1) \\
\end{equation}
\begin{equation}
\label{inter_Ham}
 H_{\mathrm{inter}}=\sum_{\mu\neq\mu',ij,\sigma}t^{\perp}_{ij}
(\hat{c}_{\mu i\sigma}^\dagger \hat{c}_{\mu' j\sigma}^{}+\hat{c}_{\mu' j\sigma}^\dagger \hat{c}_{\mu i\sigma}^{})+ 
\frac{1}{2}\sum_{\mu\neq\mu',ij} V_{ij}^{\perp}(\hat{n}_{\mu i}-1)(\hat{n}_{\mu' j}-1)\,,\hspace{0.5in}
\end{equation}
\noindent where $H_{\mathrm{intra}}$ and $H_{\mathrm{inter}}$ describe intra- and intermolecular interactions, $\hat{c}^{\dagger}_{\mu i\sigma}$ creates a $\pi$-electron of spin $\sigma$ on carbon (C) atom $i$ belonging to molecule $\mu$,
$\hat{n}_{\mu i\sigma} = \hat{c}^{\dagger}_{\mu i\sigma}\hat{c}_{\mu i\sigma}^{}$ is the number of electrons of spin $\sigma$ on atom $i$ within chromophore $\mu$,
and $\hat{n}_{\mu i}=\sum_{\sigma} \hat{n}_{\mu i\sigma}$.
The intramolecular hopping integrals $t_{ij}^{\mu}$ are between nearest neighbor C atoms $i$ and $j$,
$U$ is the Coulomb repulsion between two electrons occupying the same atomic $p_z$ orbital, $V_{ij}$ and $V_{ij}^{\perp}$ are the long range interatomic
intra- and intermolecular Coulomb interactions, respectively.
We choose $t_{ij}^{\mu}=-2.4 (-2.2)$ eV for the peripheral (internal) bonds of pentacene \cite{Aryanpour15a}, and obtain the
intermolecular hopping integrals between C atoms $i$ and $j$
separated by distance $d_{ij}$
by adjusting $\beta$ in the expression \cite{Uryu04a,Aryanpour15a}
\begin{equation}
\label{inter_hop}
t^{\perp}_{ij}=\beta~\mathrm{exp}[(d_{min}-d_{ij})/\delta]\,,
\end{equation}
where $d_{min}$ is 
taken to be the sum of van der Waals radii of two C atoms,
and  $\delta=0.045$ nm \cite{Uryu04a}.
We choose $\beta=-0.2$ to $-0.3$ eV.
With these $\beta$, 
t$_{ij}^{\perp}$ beyond the
nearest neighbor ($-$0.13 to $-0.195$ eV) and next nearest neighbor ($-$0.07 to $-0.105$ eV)
are insignificant (see Supplementary Information).

The intramolecular intersite Coulomb parameters are obtained from the expression
$V_{ij}=U/\kappa\sqrt{1+0.6117 R_{ij}^2}$, where $R_{ij}$ is the distance in $\mathring{\textrm{A}}$
between C atoms $i$ and $j$ and $\kappa$ is an effective dielectric constant \cite{Chandross97a}.
$U=11.26$ eV and $\kappa=1$ correspond to the Ohno parameters for the PPP model \cite{Ohno64a}. Nearly quantitative fits to the optical
absorption spectra of a large number of $\pi$-conjugated systems have been obtained with smaller $U$ (8.0 eV) and larger $\kappa$
(2.0) \cite{Chandross97a,Zhao06b}. More recent theoretical works suggest smaller \cite{Barcza13a} $U$ and  $\kappa$.
We have performed calculations 
for the narrow range of parameters 
$U=6.0-8.0$ eV and $\kappa=1.3-2.0$.
We have used the same functional form for $V_{ij}^{\perp}$ with the same $\kappa$ as for $V_{ij}$
\cite{Aryanpour15a}.
Excellent fits to molecular singlet and triplet energies, as well as to the 
ground state absorption spectrum for the
herringbone film are obtained with our
parametrizations of H$_{intra}$ and H$_{inter}$ (see below). 
\vskip 1pc
{\noindent \bf Results}
\vskip 0.5pc
\noindent {\bf Monomer singlet and triplet excited states.} 
The best fits to S$_1$ and T$_1$ energies are with $U=6.0$ eV, $\kappa=1.8$. For this parameter set
our calculated E(S$_1$) = 2.09 eV and E(T$_1$) =0.93 eV, to be compared against experimental E(S$_1$) = 2.15 eV \cite{Hellner72a}
and E(T$_1$) = 0.86 - 0.95 eV \cite{Burgos77a,Nijegorodov97a}.
 The monomer T$_1 \to$ T$_3$ excitation energy is 2.46 eV \cite{Hellner72a}.
The calculated T$_1 \to$ T$_3$ energy is 2.14 eV for $U=6.0$ eV, $\kappa=1.8$. From trial and error we found T$_1 \to$ T$_3$ excitation energy of
2.46 eV for $U=7.7$ eV and $\kappa=1.3$. In the following we therefore report singlet ESA spectra for $U=6.0$ eV, $\kappa=1.8$,
and the triplet and triplet-triplet ESA spectra for $U=7.7$ eV and $\kappa=1.3$. This is only for comparisons to the
experimental PA spectra. As we show in the Supplementary Information, the difference in the  
ESA spectra between the two sets of parameters is small, with $U=6.0$ eV, $\kappa=1.8$ giving slightly redshifted triplet ESAs.
\vskip 0.5pc
\noindent {\bf The herringbone structure: intra- versus intermolecular excitations.}
\vskip 0.5pc
\noindent \underbar {Singlet ground and excited state absorptions.} In Fig.~1(a) we have shown the herringbone structure for the pentacene
crystal. Our calculations are for the dimer of molecules indicated in the figure, where we have defined our axes, intermolecular
separation and dihedral angle \cite{Robertson61b,Schiefer07a}. The very weak coupling between next nearest neighbor molecules
has no effect on the calculated absorption spectrum \cite{Aryanpour15a}.
In Fig.~1(b) we have shown our calculated ground state absorption spectra for $\beta=-0.2$ and $-0.3$ eV for
the dimer of Fig.~1(a) superimposed on the low temperature experimental absorption spectrum \cite{Wilson13b}. 
The calculated spectra are obtained from diagonalization of the MRSDCI Hamiltonian matrices alone, and
no adjustment of energies were done. 
The agreement between the calculated and the
experimental absorption spectra are excellent. We are able to reproduce the Davydov splitting 
almost quantitatively; we show in the Supplementary Information that $|\beta|$ $>$ 0.3 eV is unrealistic. Our use of the purely electronic Hamiltonian does not imply absence of 
vibronic coupling \cite{Spano15a}, but indicates that the dominant interactions have been largely included in Eq.\ref{PPP_Ham}.

Our basis functions are products of many-electron
configurations on the individual pentacene molecules that constitute the dimer, within the molecular orbital (MO) description (see Methods)
\cite{Aryanpour15a,Psiachos09a}.
Each eigenstate is a quantum-mechanical superposition of these product configurations, which can not only be classfied as single, double, etc.
excitations, but also as intramolecular Frenkel, intermolecular CT or TT. 
This allows physical characterizations of all eigenstates 
in terms of their most
dominant configurations.
In Fig.~1(c) we have shown the dominant many-electron configurations and their coefficients in the normalized wavefunctions
for the three final states of the ground state absorption spectrum of Fig.~1(b) for $\beta=-0.2$ eV 
(the wavefunctions are
almost indistinguishable for $\beta=-0.3$ eV, see Supplementary Information), as well as the final states of singlet ESA
from the lowest of the three. We have labeled the optical singlet excitons as S$_0$S$_1$ for brevity, but
they are actually superpositions of S$_0$S$_1$ and S$_1$S$_0$. We have also adopted
the same convention for all other dimer excited states.
Here and below,
we retain only configurations with coefficients larger than 0.14 (terms with smaller coefficients are included for special cases). It is emphasized that although the dominant terms involve only single and double excitations
across the frontier MOs,
the complete wavefunctions have components up to quadruple
excitations and across MOs removed far from the chemical potential. The last column in Fig.~1(b) gives the extent of CT character 
of the complete wavefunction.

As seen in Fig.~1(c), the two S$_0$S$_1$ excitations at the lowest and highest energies have strong CT components, while the relatively weaker
absorption at intermediate energy is predominantly Frenkel in character.
The determination of the lowest S$_0$S$_1$ as $\sim 50$\% CT agrees with previous theoretical work \cite{Beljonne13a,Aryanpour15a}.

Experimentally, singlet PA is seen both in the visible \cite{Jundt95a,Wilson13b} and in the IR \cite{Thorsmolle09a,Walker13a,Pensack16a}.
In Fig.~1(d) we have shown the calculated ESA from the lowest S$_0$S$_1$ to S$_0$S$_2$ and S$_0$S$_3$, both in the IR. The inset shows
the calculated ESA from S$_1$ in the monomer (see inset Fig.~1(d)). We find two distinct ESAs in the IR in the herringbone structure,  
in agreement with 
experiments on TIPS-pentacene in solution \cite{Walker13a} and in a related compound in the solid state \cite{Pensack16a}, 
but in contrast to the single ESA in the IR in the monomer. 
As seen from their wavefunctions 
ESAs to S$_0$S$_2$ and S$_0$S$_3$ result from dipole-allowed excitations of the Frenkel and CT components of S$_0$S$_1$, respectively. 
The absence of a CT component to S$_0$S$_1$ in the monomer precludes the 
PA to S$_{0}$S$_{3}$. 
The nearly equal strengths of the two ESAs in the IR are due to
the nearly equal contributions by the Frenkel and CT components to the lowest S$_0$S$_1$. 
\vskip 0.5pc
\noindent \underbar {Triplet and triplet-triplet wavefunctions and ESAs.} 
We write the triplet
states also as products of dimer states. Fig.~2(a) shows our calculated triplet ESA from S$_0$T$_1$
within the herringbone structure. 
The strong ESA in the
visible occurs at a wavelength close to where monomer triplet PA is observed \cite{Hellner72a}. 
The very weak strengths of the ESAs in the NIR disagrees with the assignment of experimental PA in this region to the triplet exciton \cite{Wilson13b}. 
Fig.~2(c) gives the dimer wavefunctions for the triplet and triplet-triplet states, corresponding to the initial and final states of
the ESAs, for the same Coulomb parameters and $\beta$ as in Fig.~1(c) for consistency. The differences in the wavefunctions for the two
sets of parameters, $U=6.0$ eV, $\kappa=1.8$ and $U=7.7$ eV, $\kappa=1.3$ are small (see Supplementary Information).
The very small intensity of the triplet NIR ESA is due to the nearly 100\% Frenkel character of S$_0$T$_1$, which gives the same ESA as 
the monomer (to S$_0$T$_3$) \cite{Hellner72a}. 
The S$_0$T$_2$ state, the final state of the very weak ESA in the NIR is overwhelmingly
CT in character. The completely different characters (Frenkel versus CT) of S$_0$T$_1$ and  S$_0$T$_2$ preclude 
strong dipole coupling between them (see Supplementary Information). 

In Fig.~2(b) we have shown the calculated ESA from the $^1$(TT)$_1$ for $\beta=-0.2$ eV, 
where strong ESA in the visible is accompanied by ESA of moderate strength in the NIR. Interestingly,  
the calculated ESA energies, in both visible and NIR, 
as well as their relative intensities, agree very well with the 
PAs assigned to S$_0$T$_1$ \cite{Wilson13b}. 
In analogy with the higher energy triplet states, we have labeled the final states of triplet-triplet ESA as $^1$(TT)$_2$ and $^1$(TT)$_3$,
respectively, in Fig.~2(c). The complete $^1$(TT)$_1$ 
wavefunction, in addition to containing the product state T$_1$ $\otimes$ T$_1$,
has significant contributions from CT single excitations.
and double excitations that are products of T$_1$ and a higher energy triplet.
The CT character of $^1$(TT)$_1$ is more than an order of magnitude larger than that of S$_0$T$_1$.
The partial CT character is expected within Eq.~\ref{PPP_Ham}, since the purely
T$_1$ $\otimes$ T$_1$ state is reached from the ground state by two consecutive CT processes in opposite directions \cite{Greyson10a,Beljonne13a,Chandross99a}. 
Further, for the hypothetical linear polyene with fictitiously large bond alternation it has been shown that for E(S$_1$) $\sim$ E($^1$(TT)$_1$),
the triplet-triplet state is a strong admixture of both T$_1 \otimes$ T$_1$ and CT \cite{Chandross99a,Aryanpour15b}. Thus, the CT presence in
$^1$(TT)$_1$ is simply a result of this near resonance in pentacene. 

It becomes evident that CT contributions to the intial state is necessary for absorption in the IR. 
Within the rigid herringbone structure, the highly localized S$_0$T$_1$ has nearly vanishing CT character, and hence nearly vanishing ESA in the
NIR (see Supplementary Information section IV). 
This result presents us with a conundrum.
Either the experimental transient absorption spectra \cite{Wilson13b} are indicating the formation of $^1$(TT)$_1$ only and not free triplets, 
or the triplets that absorb in the NIR are different in character from S$_0$T$_1$ of the rigid herringbone.
\vskip 1pc
\noindent {\bf Local distortion, strongly coupled dimer and ESA spectra.}

In the following we continue to assume that SF is efficient in pentacene.
Since this necessarily requires 
long lasting PA both in the visible and NIR from S$_0$T$_1$, we study the condition necessary for such triplet PA.
We have calculated the ESA for the pentacene dimer with the locally distorted structure shown in the inset of Fig.~3(a). 
We have assumed that the dimer molecules undergo rotations and translations subsequent to photoexcitation, 
giving face-to-face stacking  
as originally suggested by Marciniak {\it et al.} \cite{Marciniak09a} (see Supplementary Information for details). 
Similar assumption has been made recently to explain excimer formation in concentrated solution \cite{Walker13a}. 
We have assumed
an idealized eclipsed geometry {\it for the sake of illustration only.} Our goal is simply to demonstrate from a model calculation
that enhanced CT, driven by photoinduced
change of the lattice locally, can give ESA seen experimentally. Computational results for more realistic geometries,
also with enhanced CT, are presented in the
Supplementary Information, where we show that the order of magnitude enhancement of S$_{0}$T$_{1}$ $\rightarrow$ S$_{0}$T$_{2}$ ESA persists in these cases. 
It is conceivable that the local distortion follows the photogeneration of $^1$(TT)$_1$, rather than S$_0$S$_1$, because of the
completely bimolecular character of the former.
Theoretical calculations have suggested that the energy barrier due to lattice strain 
prevents the rotations from occurring \cite{Kuhlman10a,Zimmerman11a}. Note, however, that (a) these calculations were for the S$_0$S$_1$ state only, and 
(b) the electronic stabilization and molecular rotation
are co-operative effects; unless the calculated electronic wavefunctions had significant CT character to begin
with, calculated rotation-induced stabilization of excited states would be far too small. 
Whether the particular local distortion of the inset of  Fig.~3(a) actually occurs or not is outside the scope of the PPP Hamiltonian, and 
to an extent even irrelevant (see below).

The calculated triplet ESAs for the distorted dimer (see Fig.~3(a)) show 
increases in oscillator strength in the NIR, by an order of magnitude relative to Fig.~2(a).
Fig.~3(b) shows the calculated triplet ESAs for both $\beta=-0.2$ and $-0.3$ eV. Fig.~3(c) shows the calculated ESA from the $^1$(TT)$_1$ now, which is largely unchanged from that in
Fig.~2(b). 
The calculated ESA wavelengths compare very favorably with the PAs observed in experiments \cite{Wilson13b}.
Fig.~3(d) shows the dominant contributions to all relevant wavefunctions.
As anticipated, there is an order of magnitude increase in the CT character of S$_0$T$_1$.
The change in the final state of the NIR ESA, S$_0$T$_2$, is insignificant. 
The NIR triplet ESA has an origin different from other ESAs in this wavelength
and requires significant CT character in S$_0$T$_1$ \cite{Psiachos09a}. 
The triplet-triplet ESAs show a small relatively insignificant redshift and little change in intensities. 
\vskip 1pc
\noindent {\bf Discussions.}

In conclusion, we have presented the first theoretical calculations of ESAs from the lowest singlet and triplet excitons, and the elusive
triplet-triplet state for a dimer of pentacene molecules, for both the herringbone structure and a structure with local distortion leading to
dimer formation.
Taken together with our wavefunction analyses, they allow 
integrated explanations of apparently contradictory experimental observations.

First, the CT character of the lowest singlet exciton has been a matter of debate. While we \cite{Aryanpour15a} and others \cite{Beljonne13a}
find close to 50\% CT contribution to this excited state, other theoretical work \cite{Zimmerman11a} have found close to vanishing CT character.
This debate cannot be resolved by comparison of theory to experimental ground state absorption spectrum alone. 
In contrast, the experimental observation of {\it two} singlet PAs in the IR \cite{Walker13a,Pensack16a}, 
as opposed to a single PA expected from the monomer (see Fig.~1(d) and insert) is a convincing proof 
of the CT character of S$_0$S$_1$. 
Similarly, Pensack {\it et al.} have argued for two PAs from the $^1$(TT)$_1$
at energies very close to those assigned to the triplet exciton \cite{Pensack16a}. 
Our calculations confirm their claim. 

We now come to our most important conclusions, viz., (a) the triplet exciton in the rigid herringbone structure is highly localized 
with practically no CT contribution, and therefore ESA in the NIR is extremely weak; and, (b) lattice relaxation that enhances the 
CT contribution to S$_0$T$_1$ is essential for observable triplet PA in the NIR. 
We believe that the much broader NIR triplet PA
that is observed experimentally, relative to the width of the singlet PA \cite{Wilson13b}, gives indirect evidence for the CT character of S$_0$T$_1$.
As seen from Fig.~3(b), small changes in $\beta$ can lead to modest but visible changes in the triplet ESA energy. Since in the real
material the extent of the photoinduced enhancement of the coupling between the members of a dimer can vary substantially, especially in polycrystalline
films, we anticipate a distribution of $\beta$ and consequently large width of the triplet NIR PA. In the case of all other NIR PAs, a distribution in $\beta$
can change the relative {\it intensities} of the PAs, but not their widths (see Supplementary Information section IV).
The observation by
Herz {\it et al.}, that the triplet PA in the NIR is delayed relative to the supposed triplet PA to visible, can also be explained. We speculate that initial 
molecular distortion follows relaxation of $^1$(TT)$_1$, driven by the electronic energy that is gained 
if the CT contribution to its wavefunction is enhanced. Additional distortion can occur now as the $^1$(TT)$_1$ dissociates into two triplets, with the 
driving force now being the electronic energy gain due to the enhanced CT contribution to each triplet. The delayed PA to S$_0$T$_2$ is then
related to slow molecular motion relative to electronic transitions, while the monomerlike PA to S$_0$T$_3$ occurs 
immediately upon the dissociation of the $^1$(TT)$_1$. 
Pensack {\it et al.} have proposed two intermediate
triplet-triplet states, a
strongly bound and a partially separated pair. Whether such a scheme can be explained within our proposed
scenario of of pre- and post-distortion $^1$(TT)$_1$ is an intriguing question.
The partial CT character of the triplet exciton may explain its long diffusion length as well as unexpectedly 
small binding energy, as
indicated by the highly efficient charge generation in pentacene/C$_{60}$ bilayer \cite{Rao10a}. We finally note that our work may also explain the paucity
of materials that show efficient SF. While the thermodynamic condition E(S$_1$) $\geq 2\times$ E(T$_1$) is necessary for efficient internal conversion to $^1$(TT)$_1$,
it is not enough for the actual fission of the latter into independent triplets, which may require morphology that allows local lattice distortions 
leading to triplets with nonnegligible CT character.
We anticipate the present work to stimulate further theoretical and experimental
research along this direction.
\vskip 1pc
\noindent {\bf Methods} The MRSDCI calculations are done iteratively for each excited state.
We begin by constructing single, double, etc. excitations within the basis space of restricted HF basis MOs that are self-consistent solutions to the extended
PPP Hamiltonian. Now at the first step of the MRSDCI 
a few (N$_{ref}$) singly and doubly excited configurations that best describe the targeted excited state are selected on the basis of a trial double-CI
calculation. The second step involves the MRSDCI calculation, in which the Hamiltonian matrix consists of single and double excitations from the
N$_{ref}$  configurations themselves, thereby constructing a much larger Hamiltonian matrix, including also the most important triple and quadruple excitations. From the total number N$_{total}$ of configurations,
a new set N$_{ref}$ of most dominant single and double excitations, whose normalized coefficients in the wavefunctions are $\geq$ 0.04, 
are chosen, and the MRSDCI calculation is performed again. 
This process of updating  N$_{ref}$ and N$_{total}$ is repeated iteratively until all configurations with coefficients $\geq$ 0.04 have been included in 
N$_{ref}$ and numerical convergence has been achieved. In the Supplementary Information we have given the N$_{ref}$
and N$_{total}$ that were used to obtain our bimolecular excited states. While N$_{ref}$ is usually of the order of a hundred, N$_{total}$ exceeds a million for most dimer eigenstates. 

The optical absorptions are calculated using the standard approach, by calculating matrix elements of the transition-dipole operator $\mu$ = e$\sum_i \vec{r_i}n_i$, where
e is the electric charge, $n_i$ has the same meaning as in Eqs. 1-3, and $\vec{r_i}$ gives the location of the C-atom within the dimer. Linewidth of 0.03 eV has been assumed in all our absorption
calculations. 

The HF basis MOs we choose are products of MOs localized on individual molecules, {\it i.e.,} they are obtained by ignoring $H_{inter}$ in the initial 
self-consistent calculation for the dimer, but are taken into consideration in each subsequent MRSDCI iteration. This procedure allows us to explicitly identify each component of a
many-body eigenstate as products of charge-neutral molecular excitations, or as CT with imbalance of charges on the two molecular fragments. Furthermore, charge-neutral
excitations can be predominantly one electron - one hole, in which case they are Frenkel excitations, or two electron - two hole, which at lowest energies are triplet-triplet,
as also evidenced from their energies which are at 2$\times$E(S$_0$T$_1$). 
\vskip 1pc
\noindent {\bf Acknowledgments.} The authors acknowledge support from NSF-CHE-1151475, the University of Arizona's REN Faculty Exploratory Research Grant and from 
Arizona TRIF Photonics. The authors are grateful to Dr. A. Rao for sending the experimental data from which the experimental absorption spectrum if Fig.~1(b) was constructed.
\vskip 1pc
\noindent {\bf Additional Information.} the authors declare no competing financial interests
\vskip 1pc
\noindent {\bf Corresponding author:} S. Mazumdar, sumit@physics.arizona.edu

\newpage
\centerline{\bf Figure Captions}
\vskip 2pc
\noindent Figure 1 (a). The herringbone structure of the pentacene. The lattice parameters were taken from references \onlinecite{Robertson61b,Schiefer07a}.
All calculations are for the dimer indicated in the figure. (b) Calculated
absorption spectra, superimposed on the experimental low temperature absorption spectrum of pentacene
film (from reference \onlinecite{Wilson13b}). (c) The dominant contributions to the normalized singlet excited states of the dimer in the pentacene
unit cell, for $U=6.0$ eV, $\kappa=1.8$, $\beta=-0.2$ eV. Only the frontier orbitals HOMO and HOMO$-$1 (red), and LUMO and LUMO + 1 (blue) and their occupancies are 
shown. The bonds represent spin-singlet superpositions of single excitations. The double excitation indicated includes the total normalized
contribution by all T$_1$ $\otimes$ T$_1$ configurations. The last column gives the overall CT contributions to the wavefunctions.
The three excited states to which ground state absorption occurs are all labeled 
S$_0$S$_1$, so that the final states observed as transient PAs could be labeled S$_0$S$_2$ and S$_0$S$_3$, respectively, for the purpose
of being consistent with the triplet and triplet-triplet nomenclature. (d) Calculated ESA spectrum from the lowest S$_0$S$_1$ in the IR. 
The inset show the
calculated singlet ESA for the monomer, where a single ESA occurs in the IR. Experimental observation of two PAs in the IR is clear signature of strong CT contribution
to S$_0$S$_1$ (see text).  
\vskip 1pc
\noindent Figure 2. Calculated triplet (a), and triplet-triplet (b) ESA spectra for pentacene in the herringbone structure, for $U=7.7$ eV, $\kappa=1.3$ 
(see text). Note that the scales along the y-axes in (a) and (b) are very different; the NIR absorption in (b) is an order
of magnitude stronger than in (a) (see also Fig.~3(a)). (c) The dominant contributions to the normalized triplet and triplet-triplet wavefunctions, for $U=6.0$ eV, $\kappa=1.8$. 
The arrows represent spin triplet bonds, with
equal admixtures of S$_z = \pm 1$ and 0. Note the one-to-one correspondence between the ESA from S$_0$T$_1$ to S$_0$T$_3$, and from
$^1$(TT)$_1$ to $^1$(TT)$_3$, which occur in the visible. 
The weak NIR triplet ESA is to be anticipated from the extremely weak CT character of S$_0$T$_1$; the much stronger absorption here in (c) is due
to the order of magnitude larger CT contribution to 
$^1$(TT)$_1$. 
\vskip 1pc
\noindent Figure 3(a). Calculated triplet ESA for $U=7.7$ eV, $\kappa=1.3$ and $\beta=-0.2$ eV (green solid curve), for the distorted lattice of the inset, 
where the two molecules constituting the dimer
are assumed to undergo rotation giving a the face-to-face structure that has also brought the molecules closer (see Supplementary Information).
The triplet ESA for the herringbone structure (dashed red curve) has been included for comparison. 
Figure 3(b). Calculated triplet ESAs, for $\beta=-0.2$ as well as $-0.3$ eV.
(c) Same as in Fig.~2(b) for $\beta=-0.2$ eV, and the distorted lattice. 
(d) Same as in Figure 2(c), with the same molecular parameters, 
for the rotated structure. Note the large increase in the
CT character of S$_0$T$_1$. 
\newpage

\begin{figure}
\hspace*{-2.5cm}                                                           
   \includegraphics[angle=90,scale=0.47]{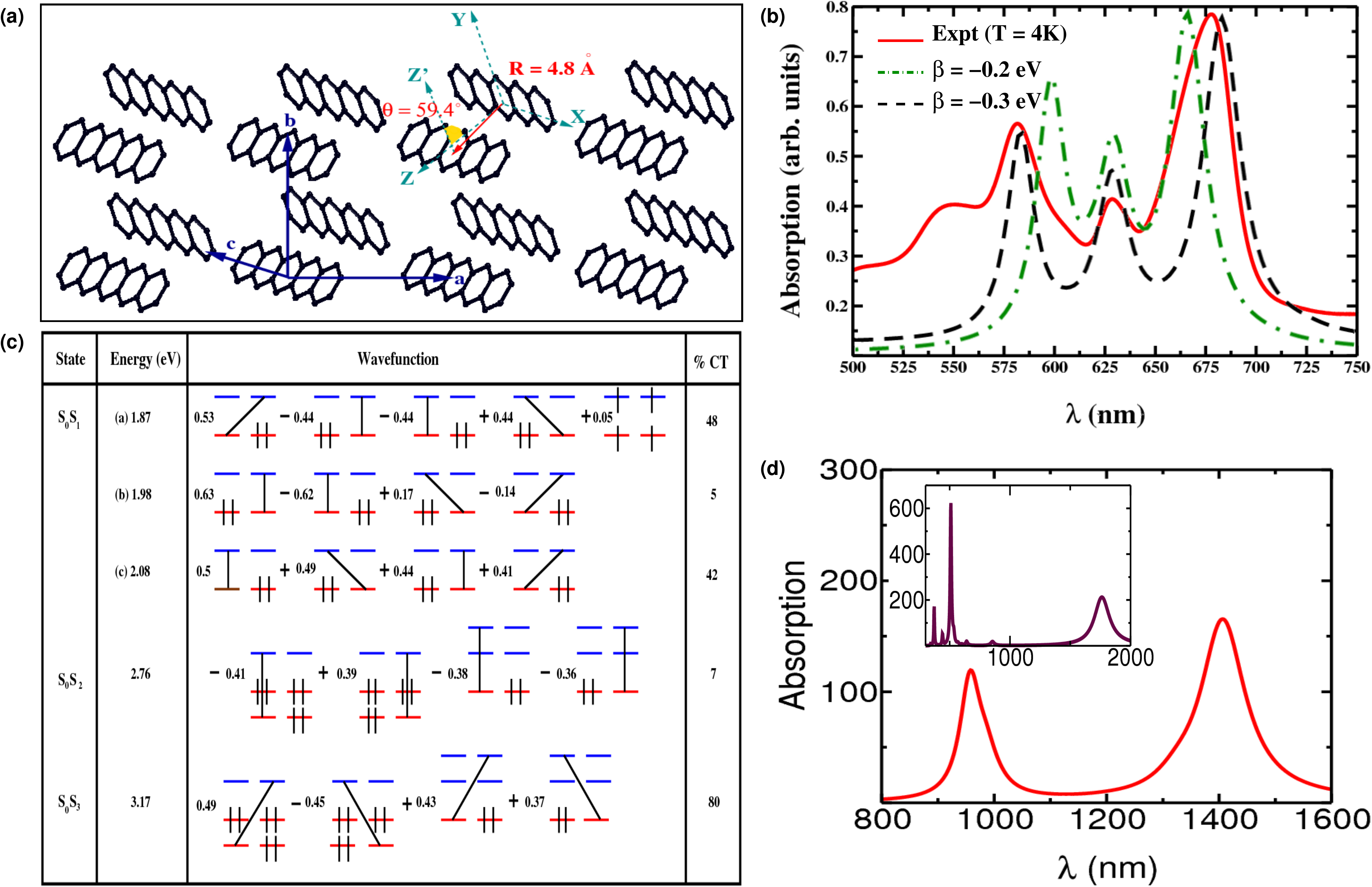}%
\caption{\textbf{Figure 1}}
\label{herringbone1}
\end{figure}

\newpage

\begin{figure}
\hspace*{-2cm}
   \includegraphics[angle=90,scale=0.5]{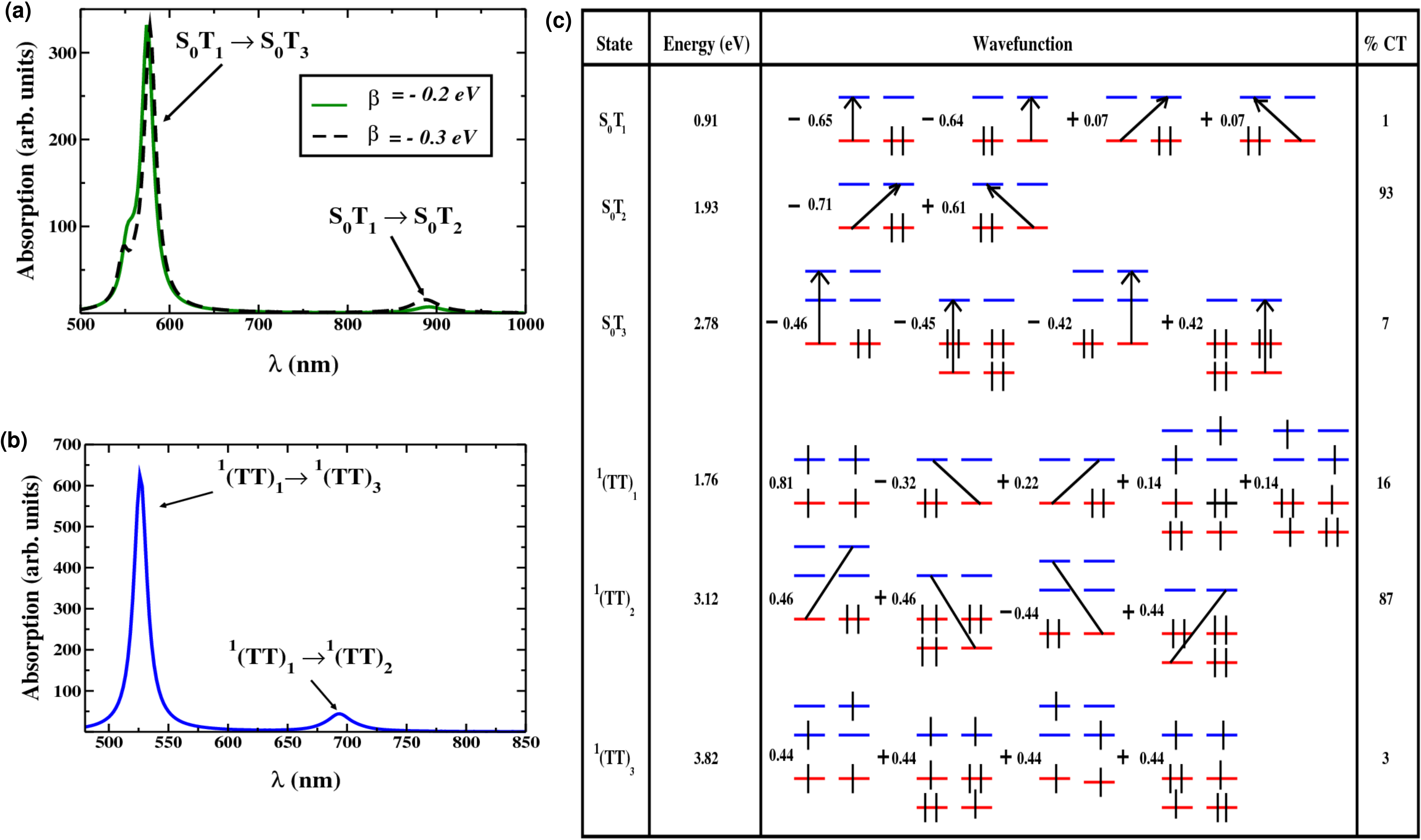}%
\caption{\textbf{Figure 2}}
\label{herringbone2}
\end{figure}

\newpage

\begin{figure}
\hspace*{-2.5cm}
   \includegraphics[angle=90,scale=0.44]{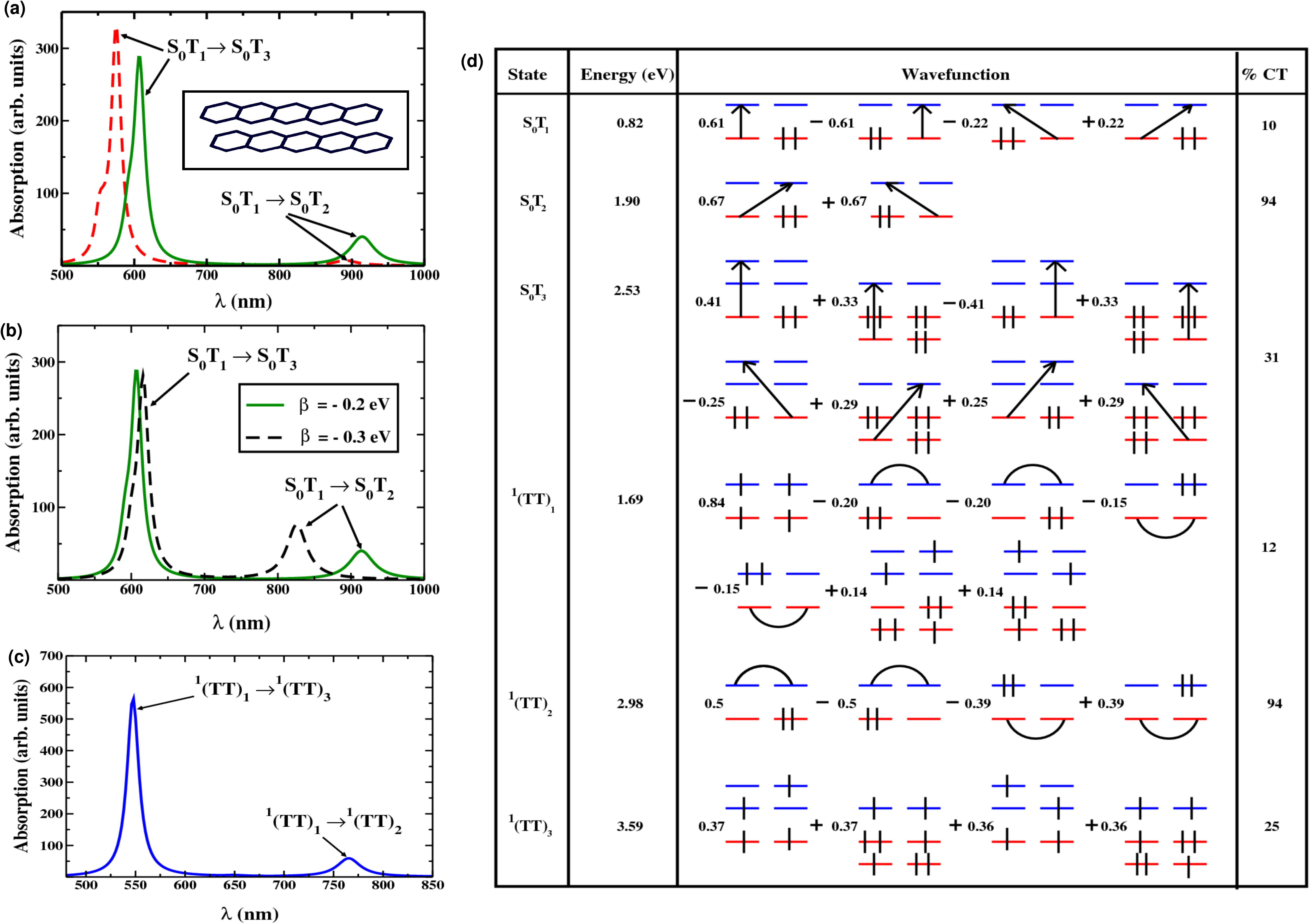}%
\caption{\textbf{Figure 3}}
\label{eclipsed}
\end{figure}



\end{document}